\documentclass[
reprint,
amsmath,amssymb,
aps,
]{revtex4-2}
\usepackage{graphicx}
\usepackage{dcolumn}
\usepackage{bm}
\usepackage{lineno}
\usepackage{xcolor}

\begin{document}
	
\preprint{APS/123-QED}
	
\title{Moir\'e-modulated $\Gamma$ valley in twisted bilayer and twisted double-bilayer MoTe$_2$}
	
\author{Wanying Chen$^{1,*}$, Hongyun Zhang$^{1,2,*,\dagger}$, Jinxi Lu$^{1,*}$, Yu Gu$^{3,*}$, Qiyun Xu$^{1,*}$, Fei Wang$^{1}$, Xuanxi Cai$^{1}$, Jiansong Li$^{1}$, Jiayong Xiao$^{3}$, Rui Chen$^{3}$, Kenji Watanabe$^{4}$, Takashi Taniguchi$^{5}$, Jose Avila$^{6}$, Pavel Dudin$^{6}$, Matthew D. Watson$^{7}$, Pu Yu$^{1}$, Shengwei Jiang$^{3}$, Wenhui Duan$^{1,8,9}$, Tingxin Li$^{3,\dagger}$, Chong Wang$^{1,9,\dagger}$, Shuyun Zhou$^{1,9,\dagger}$}
	
\affiliation{${}^{1}$State Key Laboratory of Low-Dimensional Quantum Physics and Department of Physics, Tsinghua University, Beijing 100084, People's Republic of China\\
${}^{2}$State Key Laboratory of Low-Dimensional Quantum Physics, Beijing Tsinghua Institute for Frontier Interdisciplinary Innovation, Beijing 102202, People's Republic of China\\
${}^{3}$Key Laboratory of Artificial Structures and Quantum Control (Ministry of Education), Shenyang National Laboratory for Materials Science, School of Physics and Astronomy, Shanghai Jiao Tong University, Shanghai 200240, People's Republic of China\\
${}^{4}$Research Center for Electronic and Optical Materials, National Institute for Materials Science, 1-1 Namiki, Tsukuba 305-0044, Japan\\
${}^{5}$Research Center for Materials Nanoarchitectonics, National Institute for Materials Science, 1-1 Namiki, Tsukuba 305-0044, Japan\\
${}^{6}$Synchrotron SOLEIL, L’Orme des Merisiers, Saint Aubin-BP 48, 91192 Gif sur Yvette Cedex, France\\
${}^{7}$Diamond Light Source Ltd, Harwell Science and Innovation Campus, Didcot, OX11 0DE, UK\\
${}^{8}$Institute for Advanced Study, Tsinghua University, Beijing, People’s Republic of China\\
${}^{9}$Frontier Science Center for Quantum Information, Beijing 100084, People's Republic of China}
	
\date{\today}
\begin{abstract}
Twisted MoTe$_2$ hosts intriguing correlated quantum phenomena including the fractional quantum anomalous Hall effect in twisted bilayer (t-BL) MoTe$_2$ near 3.7$^\circ$, which is sensitive to the twist angle and moir\'e superlattices. Here, we directly visualize the twist-angle-modulated electronic structure of t-BL and twisted double-bilayer (t-DBL) near this critical angle. We find that the moir\'e superlattice not only modifies the relative energy between $\Gamma$ and K valleys in t-BL MoTe$_2$, but also strongly reconstructs the $\Gamma$ valley for both t-BL and t-DBL. Specifically, the deep $p_z$-derived band at $\Gamma$ exhibits a distinct splitting that systematically varies with increasing twist angle. Theoretical analysis suggests that this modulation arises from the twist-angle-dependent  lattice relaxation, especially interfacial corrugations. Our work directly visualizes the moir\'e-modulated electronic structure and provides key spectroscopic information of lattice relaxation and interlayer interactions underlying the physics of twisted MoTe$_2$.
\end{abstract}
	
\maketitle
\begin{figure*}[htbp]
	\includegraphics[width=16.8 cm] {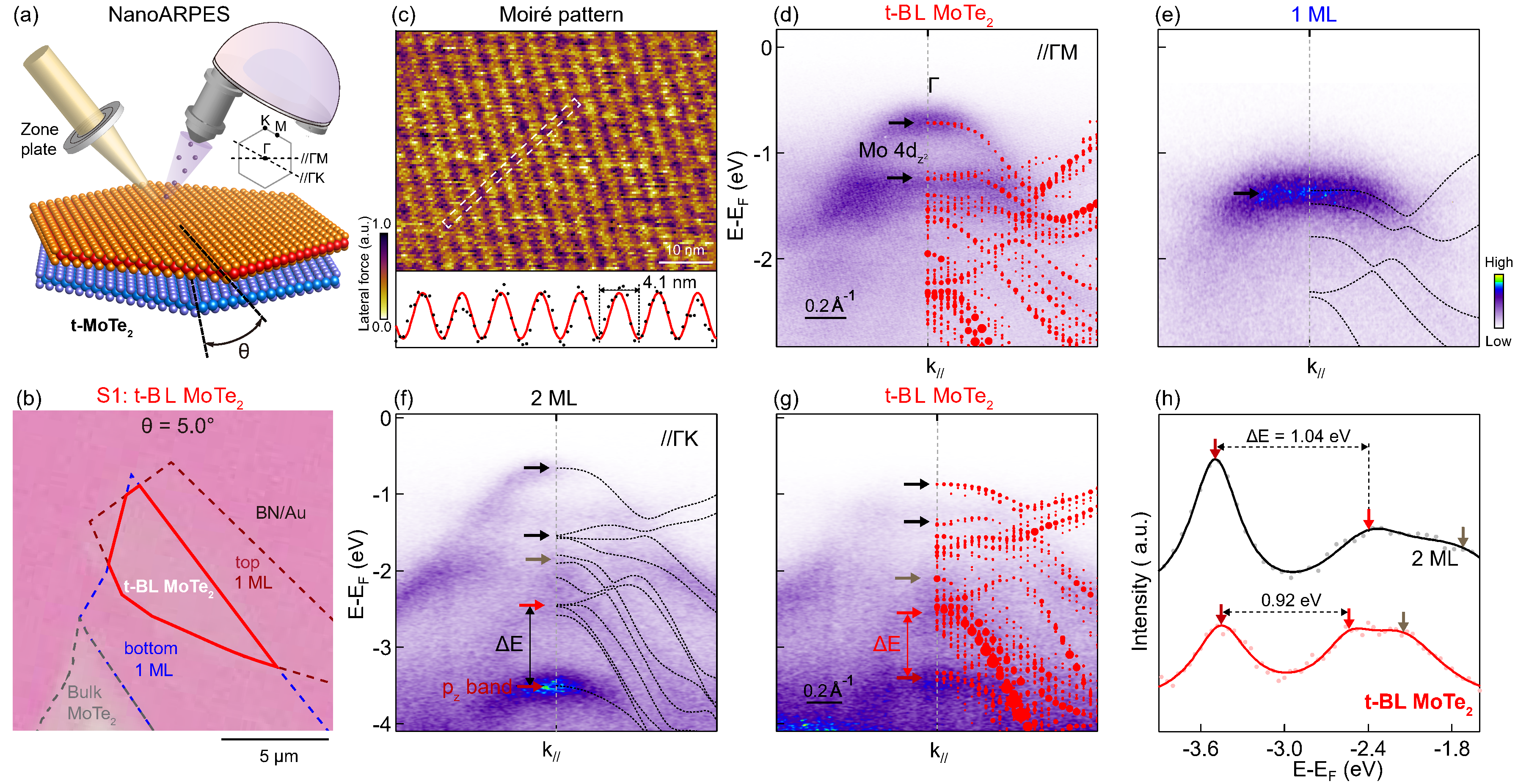}
	\caption{ {(a)} Schematic of NanoARPES measurements. {(b)} Optical image of t-BL MoTe$_2$ with $\theta = 5.0^{\circ}$ (sample S1). {(c)} L-AFM image showing a moir\'{e} period of $\lambda$  = 4.1 $\pm$ 0.3 nm. {(d),(e)} Dispersion images measured on t-BL MoTe$_2$ (sample S1) and 1 ML MoTe$_2$ along the $\Gamma$-M direction. The calculated dispersions are over-plotted for comparison. {(f),(g)} Dispersion images measured along the $\Gamma$-K direction on natural 2 ML and t-BL MoTe$_2$ (sample S2, $\theta$ = 5.0$^\circ$).  {(h)} Comparison of EDCs at the $\Gamma$ point for 2 ML and t-BL MoTe$_2$ to extract the deep energy splitting, marked by arrows in (f),(g).}
	\label{F1}
\end{figure*}

Twisted transition metal dichalcogenide (TMDC) moir\'e superlattices provide a highly tunable platform for exploring correlated and topological quantum phases \cite{Bernevig_PRX2021,Sheng_NC2011,Wen_PRL2011,Sarma_PRL2011,Christopher_PRL2011}. Among these, twisted MoTe$_2$ has attracted extensive interests due to its rich correlated phenomena, including ferromagnetism \cite{anderson2023sci,park2025np}, superconductivity \cite{Li_arXiv2025}, and fractional Chern insulators (FCIs) exhibiting fractional quantum anomalous Hall effect (FQAHE) \cite{cai2023nature,xu2023prx,zeng2023nature,park2023nature,kang2024nature,ji2024nature,redekop2024nature}. The FQAHE observed in twisted bilayer (t-BL) MoTe$_2$ distinguishes it from other TMDC moir\'e systems due to its delicately engineered topological bands \cite{reddy2023prb,xu2024pnas,xu2025np,liu2024arxiv,li2021nature,wang2025prl,devakul2021nc,wang2025nature}. The formation of these states is intimately tied to the moir\'e superlattice, which modulates the interlayer coupling. This coupling can be further tuned by layer number and stacking configuration \cite{Wu_arXiv2025}. In natural (untwisted) multilayer TMDCs, adding layers drives the valence band maximum from K toward the $\Gamma$ valley \cite{jin2013direct}, enabling valley-controlled correlated states such as valley-charge-transfer insulators and tunable metal-insulator transitions, as demonstrated by optical and transport measurements of t-BL, twisted double bilayer (t-DBL) WSe$_2$, and twisted multilayer heterostructures \cite{Brian2024np,ma2025nm,foutty2023nm,Wang2025nm}. Directly probing the modulation of the electronic structures both at the K and $\Gamma$ valleys by the moir\'e superlattice is therefore important. So far, while scanning probe techniques have provided real-space signatures of moir\'e-driven lattice relaxation \cite{Thompson2025_NatPhys_Skyrmion,liu2024arxiv} and angle-resolved photoemission spectroscopy (ARPES) measurements have revealed the shifting of top valence bands at $\Gamma$ and K valleys \cite{deng2025arxiv,chen2025arxiv} in t-BL MoTe$_2$, a direct, momentum-resolved picture of how the electronic band structure is modulated collaboratively by the twist angle and layer number has remained elusive, hindering a complete understanding of the interplay between moir\'e superlattice and emergent quantum properties.

\begin{figure}[htbp]
    \includegraphics[width=0.49\textwidth]{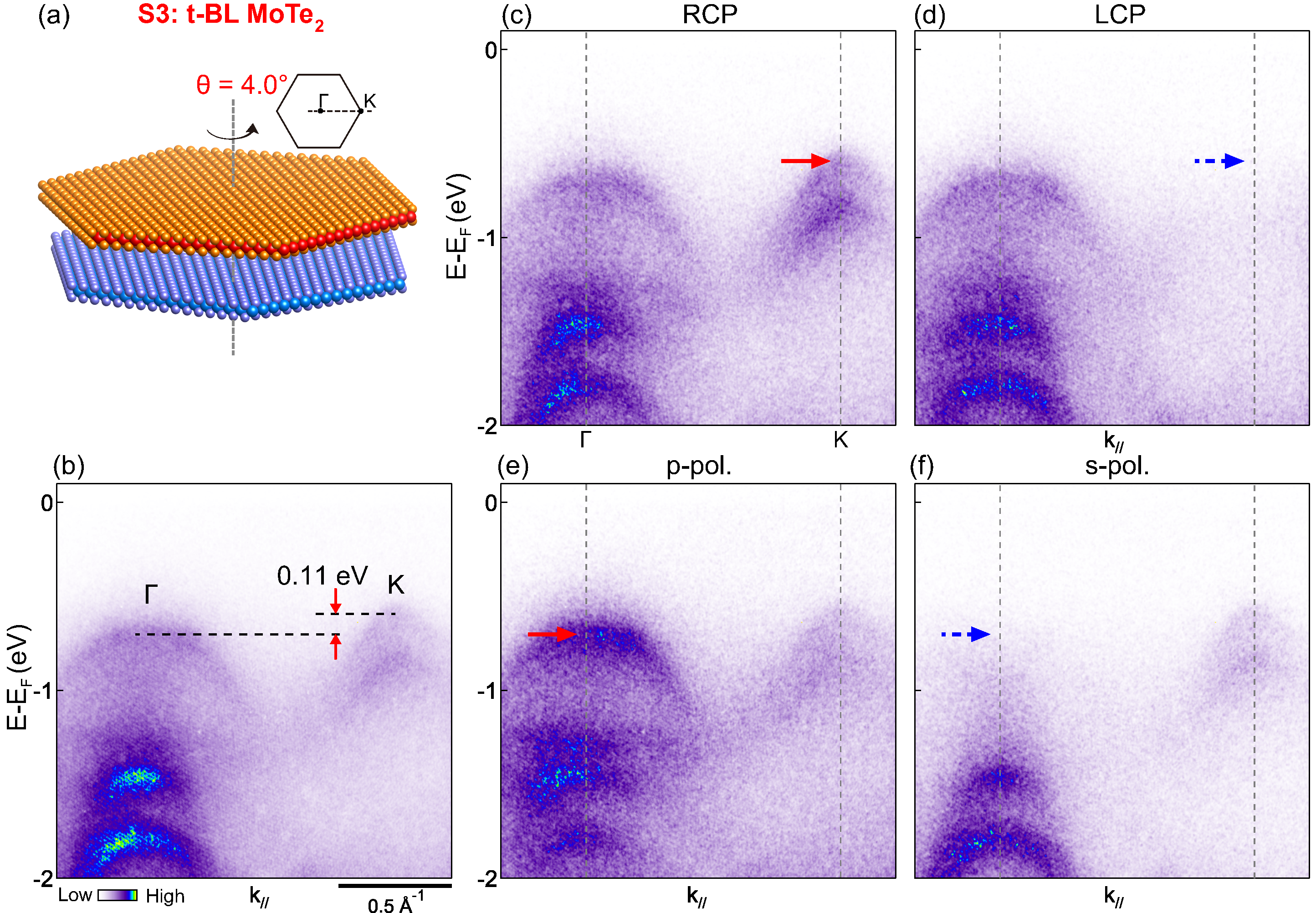}
	\caption{ {(a)} Schematic of the t-BL MoTe$_2$  with a twist angle of $\theta=4.0^{\circ}$ (sample S3). {(b)} Dispersion image measured on t-BL MoTe$_2$ along the $\Gamma$-K direction. {(c),(d)} Dispersion images acquired under right/left circular polarization (RCP/LCP). {(e),(f)} Dispersion images acquired under horizontal/vertical linear polarization ($p$-pol./$s$-pol.).}
	\label{F2}
\end{figure}

Here, by performing nano-spot ARPES [NanoARPES, Fig.~1(a)] measurements, we directly visualize the modulated electronic structures of not only t-BL, but also the hitherto-unexplored t-DBL MoTe$_2$ samples. We find that the interlayer interaction not only modifies the relative energies between $\Gamma$ and K valleys, but more importantly, strongly modifies the $p_z$-derived bands at high binding energy at the $\Gamma$ valley. The splitting of the deep valence bands therefore can serve as a direct spectroscopic measure of this interlayer coupling. Combined with theoretical calculations, our results suggest that the $\Gamma$ valley splitting originates from the moir\'e-induced lattice relaxation, which modulates the interfacial Te-Te spacing. Our work provides spectroscopic insights on the moir\'e-driven electronic structure reconstruction that underlies the emergence of novel quantum phases in twisted MoTe$_2$ platform.

\begin{figure*}[htbp]
	\includegraphics{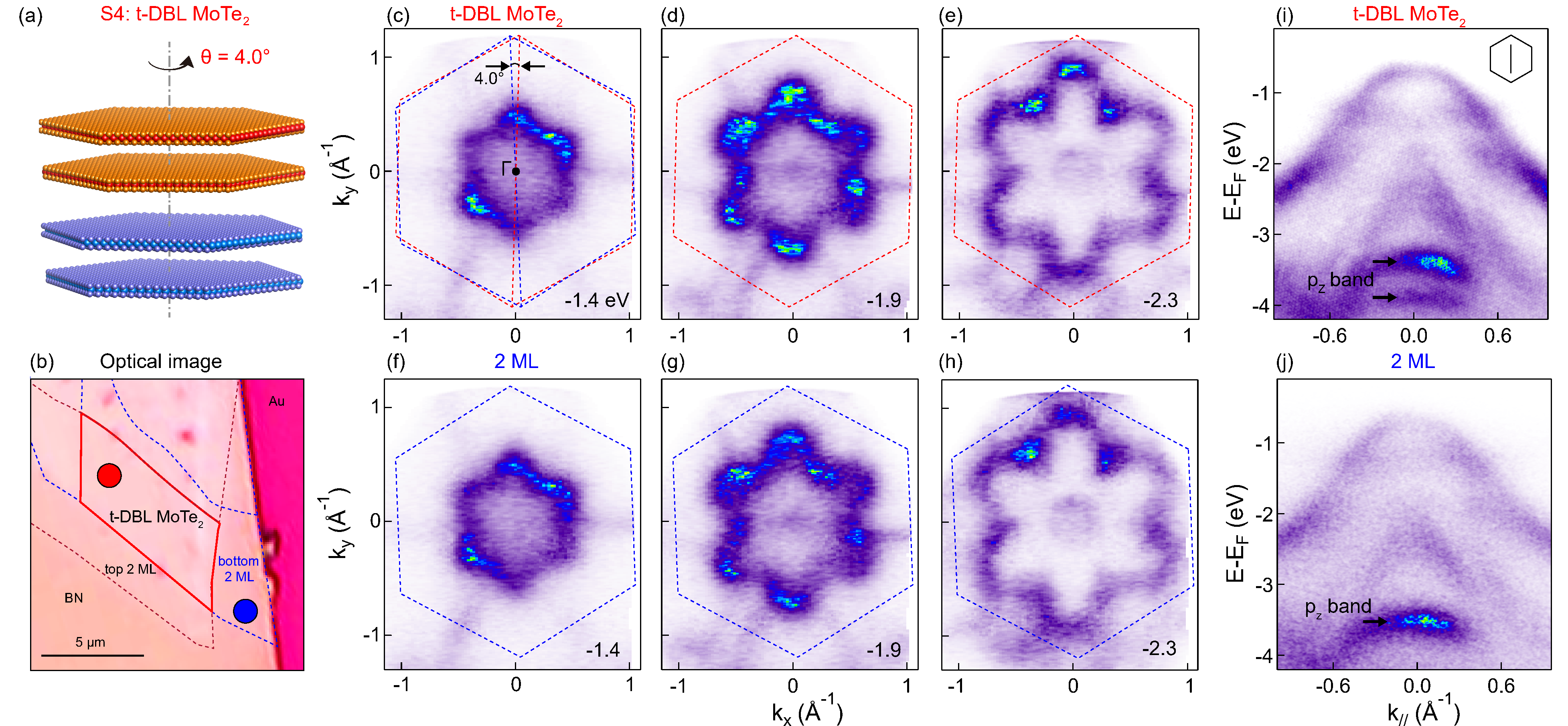}
	\caption{
	{(a)} Schematic of t-DBL MoTe$_2$ with $\theta = 4.0^{\circ}$ (sample S4). {(b)} Optical image of sample S4. {(c)-(e)} Intensity maps at  energies of -1.4, -1.9 and -2.3 eV, showing the twisted region with Brillouin zones marked by red and blue hexagons. {(f)-(h)} Intensity maps measured on the bottom 2 ML region at the same energies as (f)-(h), with its Brillouin zone marked by blue hexagon. {(i)} Dispersion image measured along the $\Gamma$-K direction on the t-DBL MoTe$_2$ region. {(j)} Dispersion image measured on bottom 2 ML along the $\Gamma$-K direction.}\label{Fig3}
\end{figure*}

High-quality t-BL and t-DBL MoTe$_2$ samples covering twist angle near 4.0$^\circ$ to 5.0$^\circ$ were fabricated using a clean, dry-transfer method \cite{wang2013transfer,cao2016prl} (see Methods and Supplementary Fig.~S1). Figure~1(b) shows an optical image of a representative t-BL MoTe$_2$ (sample S1, $\theta$ = 5.0$^\circ$), with the twisted region highlighted in red. The twist angle was confirmed by lateral-force atomic force microscope (L-AFM) measurements, as shown in Fig.~1(c) (see also Supplementary Fig.~S2). The extracted moir\'e period of  $\lambda$  = 4.1 $\pm$ 0.3 nm corresponds to a twist angle of $\theta = 5.0^{\circ}\pm0.4^\circ$. 

We find that the electronic structure at the $\Gamma$ valley is directly sensitive to the interlayer coupling. Figure~1(d)-1(e) compares NanoARPES dispersion images along the $\Gamma$-M direction for t-BL and monolayer (1 ML) samples, with calculated band structures over-plotted and showing overall agreement. A key signature of the interlayer coupling is the splitting of the valence band at the $\Gamma$ valley, primarily derived from Mo 4$d_{z^2}$ orbitals \cite{zhang2018nl}, into two distinct peaks in the t-BL sample [indicated by black arrows in Fig.~1(d)]. 
The electronic structure of t-BL MoTe$_2$ is further modulated by the moir\'e superlattice.  Figure 1(f)-1(g) shows a comparison of dispersion images between natural 2H-bilayer (2 ML, $\theta$ = 60.0$^\circ$) and t-BL MoTe$_2$ ($\theta$ = 5.0$^\circ$), with corresponding calculated band structures over-plotted. The splitting of the top valence band from the Mo 4$d_{z^2}$ orbitals is reduced in the t-BL sample compared to 2 ML sample (marked by black arrows). 
The reduced splitting is even more pronounced for the deep valence bands at deep binding energy, which are mainly contributed by the Te $p_z$ orbitals and are therefore highly sensitive to the out-of-plane interlayer interactions. Their energy splitting ($\Delta$E) at the $\Gamma$ point decreases from 1.04 $\pm$ 0.02 eV in the 2 ML to 0.92 $\pm$ 0.02 eV in the t-BL MoTe$_2$ [Fig.~1(h)], demonstrating a substantial modification of the electronic structure by the moir\'{e} superlattice. 
	
The moir\'e superlattice also shifts the relative energies between $\Gamma$ and K valleys. Figure 2(b) shows a dispersion image for t-BL MoTe$_2$ sample [S3, $\theta$ = 4.0$^\circ$, Fig.~2(a)], which is close to the twist angle where FQAHE is observed. Unlike natural 2 ML case, where the valence band maximum of the $\Gamma$ and K valleys lie at similar energy \cite{deng2025arxiv}, the K valley valence band maximum in the t-BL sample lies 0.11 eV above $\Gamma$ valley, similar to recent NanoARPES works \cite{deng2025arxiv,chen2025arxiv}. Such relative energy shift indicates a transition from indirect to direct gap semiconductor.  The distinct orbital contributions of the $\Gamma$ and K valleys also give rise to valley-contrasting polarization responses in ARPES. The K valley exhibits pronounced circular dichroism [Fig.~2(c)-2(d), Supplementary Fig.~S3], while the $\Gamma$ valley shows a strong linear dichroism [Fig.~2(e)-2(f)]. Such valley-contrasting behavior originates from the specific orbital angular momentum and Berry curvature of each valley \cite{Cho2021_SciRep_CD, Cho2018_PRL_CD,Schuler2022_PRX_CD, Schuler2020_SciAdv_CD}, which could be potentially affected by the moir\'e superlattice through its control over their relative positions.

The moir\'e modulated $\Gamma$ valley is also observed in t-DBL MoTe$_2$, formed by stacking two 2H-MoTe$_2$ bilayers with a twist angle [Fig.~3(a)-3(b)].
Constant-energy maps at the t-DBL region [Fig.~3(c)-3(e)] reveal that the main hole pocket centered at $\Gamma$ point evolves from a hexagonal shape to a flower-like pattern with increasing binding energy. A similar evolution is also observed in the bottom 2 ML region [Fig.~3(f)-3(h)], yet with its Brillouin zone rotated by the twist angle. A comparison of dispersion images along the $\Gamma$-K direction highlights the effect of moir\'e superlattice on the $\Gamma$ valley between the twisted and bottom 2 ML regions [black arrows in Fig.~3(i)-3(j)]. The most prominent change is the lower $p_z$ band [black arrow in Fig.~3(j)], which is much sharper and stronger than other bands. This band clearly splits into two in the t-DBL sample [Fig.~3(i)], demonstrating the modulation of the $\Gamma$ valley in the moir\'e superlattice.
	
To track the evolution of the $p_z$ band splitting with twist angles in t-DBL MoTe$_2$, we show in Fig.~4(a)-4(d) dispersion images at the $\Gamma$ valley on four t-DBL MoTe$_2$ samples with $\theta = 0.0^{\circ}$ (natural 4 ML sample, with AB-AB stacking), $4.0^{\circ}$, $10.0^{\circ}$ and $63.0^{\circ}$ (sample S5 and S6, see also Supplementary Fig.~S4 for more data). While the low-energy bands remain largely unchanged, the deep $p_z$-derived bands exhibit pronounced modifications. The energy splitting of these bands decreases with the twist angle and increases at $\theta$ = $63.0^{\circ}$, as seen in the dispersion images [dotted curves in Fig.~4(a)-4(d)] and summarized schematically in Fig.~4(e)-4(h). By fitting energy distribution curves (EDCs) at the $\Gamma$ point [Fig.~4(i)], we extract the energy splitting $\Delta E$, which decreases from 586 $\pm$ 10 meV at $0.0^{\circ}$ to 493 $\pm$ 15 meV at $10.0^{\circ}$, and increases to 630 $\pm$ 20 meV at $63.0^{\circ}$  [Fig.~4(j)]. Such evolution provides a direct, quantitative measure of how the twist angle modulates the  interlayer coupling in t-DBL MoTe$_2$, whose origin will be discussed below.

\begin{figure*}[htbp]
	\includegraphics[width=16.8 cm]{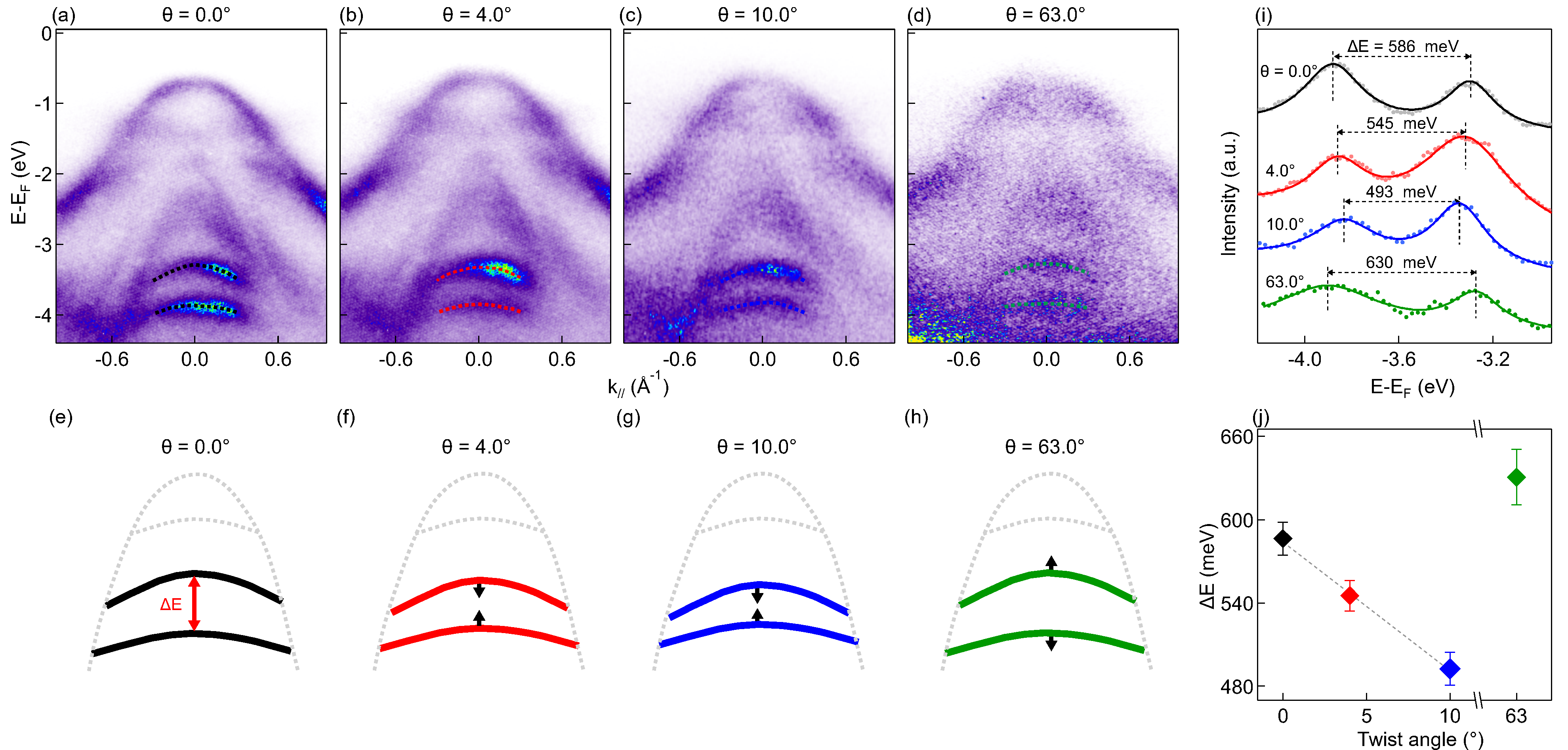}
	\caption{
		{(a)-(d)} Dispersion images measured on t-DBL MoTe$_2$ with $\theta = 0.0^{\circ}$, $4.0^{\circ}$, $10.0^{\circ}$, $63.0^{\circ}$ along the $\Gamma$-K direction.  {(e)-(h)} Schematic summary of band structures at deep energies for different twist angles.  {(i)} EDCs at the $\Gamma$ point to extract energy splitting of the lower $p_z$ band at different twist angles. {(j)} Extracted energy splitting of the deep valence bands as a function of twist angle.
	}\label{Fig4}
\end{figure*}

To elucidate how the twist angle modulates  the interlayer interaction and  the $\Gamma$ valley electronic structure, we performed first-principles density functional theory (DFT) calculations combined with DFT-trained machine-learning interatomic potentials (MLIPs) \cite{RN1697} and band-unfolding analysis (see Supplementary Information for details). Figure~5(a)–5(d) shows the calculated moir\'e bands for $\theta$ = $0^\circ$, $3.89^\circ$, $9.43^\circ$, and $60^\circ$ respectively. The calculated band splittings (marked by blue arrows) follow the same trend as our experimental observations, confirming the twist-angle-dependent modulation of the $\Gamma$ valley. 

The modulation of the band splitting arises from changes in the strength of the interlayer coupling, as quantified by the calculated average distance between the interfacial Te-Te layers [Fig.~5(e)–5(j)].
The natural 4 ML sample ($\theta = 0^\circ$) with AB-AB stacking [Fig.~5(e)] exhibits the shortest interfacial Te–Te spacing of 3.21~\AA{} and is the most stable configuration (H stacking). The resulting strong interlayer hybridization produces the maximum splitting of the $p_z$-derived band. When a twisted moir\'e structure is formed, the moir\'e superlattice initially samples an approximately equal distribution of all possible local stackings, including energetically unfavorable ones. These high-energy stacking configurations generally have a larger interfacial Te-Te spacing with spatially modulated corrugations [Fig.~5(g)-5(h)], which leads to a reduced splitting of the $p_z$-derived band.
The relaxation of the moir\'e structure partially compensates this increase in the interfacial Te-Te spacing, because the fractional area occupied by the most stable stacking, with the shortest Te-Te spacing, grows upon relaxation. Moreover, smaller twist angles favor stronger moir\'e relaxation, so the $4^\circ$ superlattice undergoes more pronounced reconstruction and therefore attains a smaller average interfacial Te-Te spacing (3.51~\AA{}) than the $10^\circ$ superlattice (3.56~\AA{}). 
\begin{figure*}[htbp]
	\includegraphics{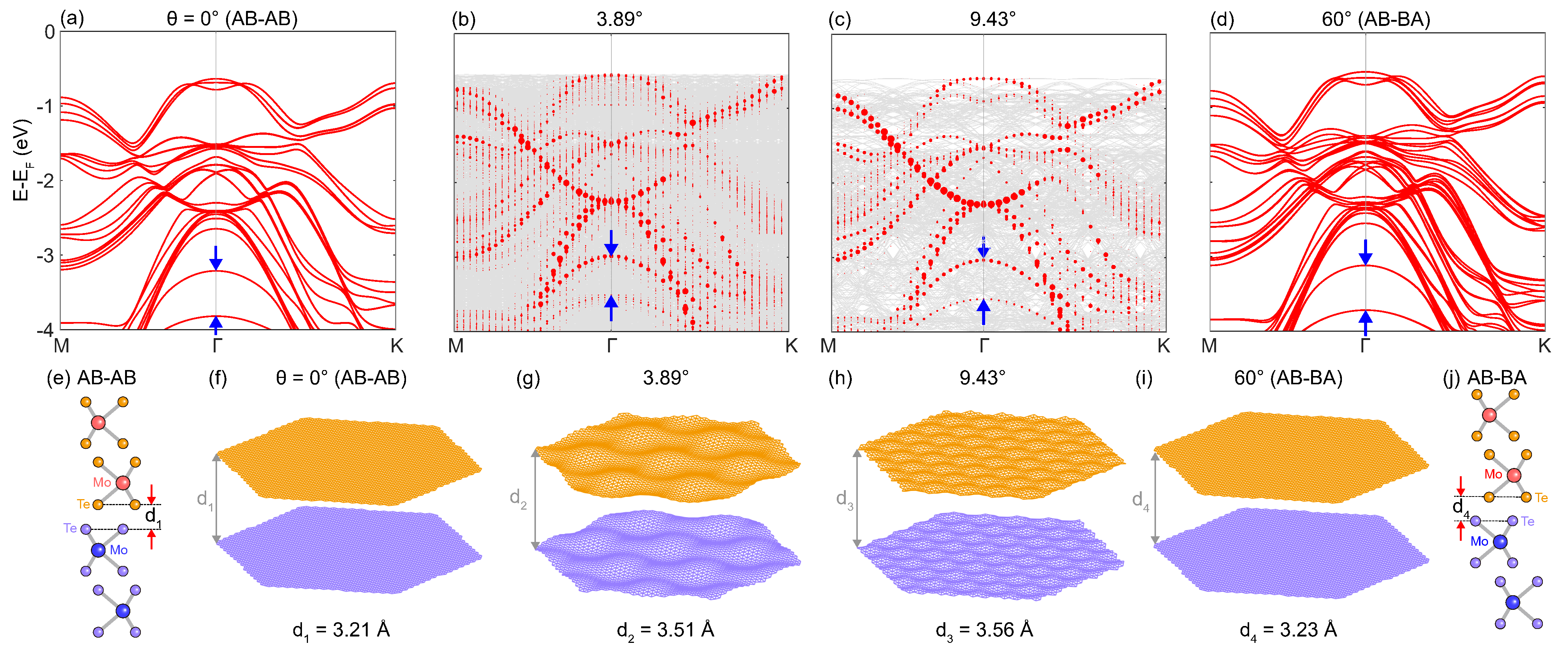}
	\caption{
		{(a)-(d)} Calculated band structure for t-DBL MoTe$_2$ with twist angles of $\theta = 0^{\circ}$, $3.89^{\circ}$, $9.43^{\circ}$, $60^{\circ}$. {(e)} Atomic structure of AB-AB stacked MoTe$_2$. {(f)-(i)} Lattice relaxation with modulated effective interlayer spacing for different twist angles. {(j)} Atomic structure of AB-BA MoTe$_2$, which corresponds to 60$^\circ$ t-DBL MoTe$_2$ with rhombohedral stacking.	}\label{Fig5}
\end{figure*}			
At $\theta = 60^\circ$, with AB-BA stacking [Fig.~5(j)], the structure adopts a metastable rhombohedral stacking (the R phase), accompanied by a reduction of the interlayer spacing to 3.23~\AA{} [Fig.~5(i)], slightly larger than that at $\theta = 0^\circ$ (H stacking). The resulting strong interlayer coupling again leads to a larger $p_z$-band splitting. Such R-stacking MoTe$_2$ has been proposed to host nontrivial topology~\cite{Zhang2024_NatComm_Skyrmion,Wu_arXiv2025,fan2025arxiv}, making it a compelling system for future studies. Although the structural variations from the calculations are too subtle to be directly resolved experimentally, our analysis of the deep valence bands at the $\Gamma$ valley provides a direct window into how the twist angle modulates lattice relaxations. This strategy, previously applied to twisted bilayer graphene~\cite{Zhou_NM2024}, offers a general route for probing interlayer coupling and atomic reconstruction in a wide range of twisted bilayers and van der Waals heterostructures.

In summary, we have directly visualized the twist-angle-dependent electronic structure of twisted MoTe$_2$. We establish the energy splitting of $p_z$-derived bands at the $\Gamma$-valley as a quantitative spectroscopic measure of interlayer coupling. Our combined experimental and theoretical analysis reveals that this coupling is governed by moir\'e-driven lattice relaxation, which modulates the effective interlayer spacing with interfacial Te-Te corrugations. These results provide a direct spectroscopic link between atomic-scale structure and electronic properties, offering fundamental insights for engineering quantum phases in moir\'e materials.
\begin{acknowledgments}
This work is supported by the National Key R$\&$D Program of China (Grant No.~2021YFA1400100), the National Natural Science Foundation of China (Grant No.~12522403), Tsinghua University Initiative Scientific Research Program (Grant No.~20251080106), National Natural Science Foundation of China (No.~12421004, 12234011, 92250305), and the New Cornerstone Science Foundation through the XPLORER PRIZE. K.W. and T.T. acknowledge support from the JSPS KAKENHI (Grant Numbers 21H05233 and 23H02052), the CREST (JPMJCR24A5), JST and World Premier International Research Center Initiative (WPI), MEXT, Japan. We acknowledge SOLEIL for the provision of synchrotron radiation facilities at the beamline ANTARES, the Diamond Light Source for the provision of synchrotron radiation facilities at beamline I05, and the Beamline 07U of the Shanghai Synchrotron Radiation Facility.
\\\noindent$^\ast$These authors contributed equally to this work.
$^\dagger$Corresponding author: 
\\zhanghy.2015@tsinghua.org.cn; 
\\ txli89@sjtu.edu.cn; 
\\ chongwang@mail.tsinghua.edu.cn; 
\\syzhou@mail.tsinghua.edu.cn
\end{acknowledgments}

%

\end{document}